\pdfoutput=1
\documentclass[runningheads]{llncs}
\usepackage{graphicx}
\usepackage[dvipsnames]{xcolor}
\usepackage{booktabs}
\usepackage{algorithm}
\usepackage{algpseudocode}
\usepackage{algorithmicx}
\usepackage{amsmath}
\usepackage{amssymb}
\usepackage{amsfonts}
\usepackage{multirow}

\DeclareMathOperator*{\argmax}{argmax}
\DeclareMathOperator*{\argmin}{argmin}
\newcommand{\point}[1]{\par\smallskip\noindent\textbf{#1.}}

\begin{document}

\title{Step on the Gas? A Better Approach for Recommending the Ethereum Gas Price}

\titlerunning{Step on the Gas?}

\author{Sam M. Werner\thanks{The author would like to thank the \textit{Brevan Howard Centre for Financial Analysis} for its financial support.} \and
Paul J. Pritz \and Daniel Perez\thanks{The author would like to thank the \textit{Tezos Foundation} for its financial support.}
}

%%% Note: If we have more "thanks" to mention, we can simply add an acknowledgement section in the end.

\authorrunning{S.M. Werner et al.}
\institute{Imperial College London, United Kingdom \\
\email{\{sam.werner16,paul.pritz18,daniel.perez\}@imperial.ac.uk}}

\maketitle

\begin{abstract}
\label{ref:abstract}
In the Ethereum network, miners are incentivized to include transactions in a block depending on the gas price specified by the sender.
The sender of a transaction therefore faces a trade-off between timely inclusion and cost of his transaction.
Existing recommendation mechanisms aggregate recent gas price data on a per-block basis to suggest a gas price.

We perform an empirical analysis of historic block data to motivate the use of a predictive model for gas price recommendation.
Subsequently, we propose a novel mechanism that combines a deep-learning based price forecasting model as well as an algorithm parameterized by a user-specific urgency value to recommend gas prices. 

In a comprehensive evaluation on real-world data, we show that our approach results on average in costs savings of more than 50\% while only incurring an inclusion delay of 1.3 blocks, when compared to the gas price recommendation mechanism of the most widely used Ethereum client.
\end{abstract}

% --- Introduction ---
\section{Introduction}
\label{ref:introduction}
Since the introduction of Ethereum~\cite{Buterin2014} and it's virtual machine, participants have been able to create so-called \textit{smart contracts}, i.e. programs that encapsulate the logic for governing funds.
As these contracts have to be executed by all participating nodes in the Ethereum network, the sender of a transaction has to pay for the computational cost of execution in units of \textit{gas}.
The amount of gas to be paid by the sender of a transaction depends on the complexity of executing a smart contract's logic.
Additionally, the sender is required to specify the \textit{gas price}, which he will have to pay per unit of consumed gas.
The product of the gas cost and price determines the transaction fee, which is received by the miner who includes the transaction in a block.
Hence, setting an appropriate gas price is critical for having a transaction included in a timely manner.
While Ethereum employs a hard coded and transparent gas cost model, there does not exist any embedded mechanism for computing how much a sender of a transaction should pay per unit of gas.
The gas price is instead determined by the supply and demand for computational resources.
Therefore, choosing an optimal gas price can be challenging, as underpaying likely results in a transaction not being included by miners, whereas overpaying leads to avoidable costs.

The most widely used gas price prediction mechanism is implemented by the popular Ethereum client Geth~\cite{github:geth}.
This and comparable mechanisms only use recent gas prices and merely aggregate past data to heuristically recommend the gas price for a transaction.

In this paper, we present a novel approach for gas price prediction, motivated by the empirical analysis of a period of 522,213 blocks.
We find significant seasonality in the gas price data, suggesting that this can be predicted using a machine learning model.
We propose the use of Gated Recurrent Units \cite{GRU_paper_2015} as these have been shown to be suitable for capturing such patterns.
Consequently, we design an algorithm for choosing the gas price for a transaction, which leverages the predictions of our model while allowing to specify the transaction's urgency.
Our evaluation on real-world data shows that the proposed approach significantly outperforms the most widely-used Ethereum client Geth~\cite{ethernodes}.

\point{Contributions}
Our contributions are as follows:
\begin{enumerate}
    \item We present a comprehensive empirical analysis of the Ethereum gas price over a period of three months and identify seasonal patterns in the data,
    \item We propose a deep-learning based model to predict the gas price and combine this with a novel algorithm for recommending the gas price for a transaction,
    \item We evaluate our model on real-world data and show that it outperforms the most widely used gas price recommendation approach, resulting on average in costs savings of more than 50\% while only incurring an inclusion delay of 1.3 blocks compared to Geth.
\end{enumerate}
\point{Structure}
The remainder of this paper is organized as follows.
Section~\ref{ref:background} introduces the background of Ethereum and its embedded gas mechanism.
An empirical analysis of Ethereum gas prices is presented in Section~\ref{ref:empirical-analysis}.
We propose a methodology for better gas price recommendation in Section~\ref{ref:methodology}, before evaluating our model's results in Section~\ref{ref:results}.
Related work is discussed in Section~\ref{ref:related_work}.
Lastly, we conclude in Section~\ref{ref:conclusion}.

% --- Background ---
\section{Background}
\label{ref:background}

In this section, we first provide a brief overview of the workings of the Ethereum network. 
Subsequently, we examine in greater detail the gas cost and pricing mechanisms used in Ethereum.

\subsection{Ethereum} 
Ethereum employs a Proof-of-Work consensus mechanism as first introduced by Bitcoin~\cite{nakamoto2008bitcoin}.
In such a protocol, transactions are grouped into blocks and Ethereum's block arrival time is approximately 13 seconds~\cite{etherscan:block_time}.
Ethereum allows for the creation of so-called \textit{smart contracts}.
These are programs which define a set of rules using a Turing-complete programming language, typically Solidity~\cite{Dannen:2017:IES:3103305}, that can be invoked by network participants.
An Ethereum account balance is expressed in the underlying currency Ether (ETH) and directly altered via state transitions caused by transactions.
The consensus rules governing transaction validity are implemented by the \textit{Ethereum Virtual Machine} (EVM), a low-level stack machine which executes the compiled EVM bytecode of the smart contract.
Operations performed by the EVM consume \textit{gas}, a virtual unit of account used to measure the computational cost of executing a transaction. 
By design, each EVM instruction has a hard-coded\footnote{Note that via a hard-fork, the Ethereum Improvement Proposal 150~\cite{erc150} re-aligned gas costs for instructions involving I/O-heavy operations.} gas cost~\cite{wood2014ethereum}.
The total execution cost has to be paid for by the sender of a transaction.

\subsection{Gas Mechanism}
The total execution cost for a contract consists of two components, namely the gas cost in units and gas price per unit.
The gas cost is split into a fixed base cost of 21\,000 gas and an execution cost dependent on the instructions executed while running the contract.

\point{Gas Limit}
Due to the Turing-completeness of the EVM, the exact computational cost of a transaction cannot be predetermined.
Hence, the sender is required to specify a \textit{gas limit}, or the maximum amount of gas that may be consumed.
As the computational steps of a transaction are executed, the required gas is subtracted from the paid gas.
Once a transaction is completed, any unused gas will be refunded to the sender.
Should a transaction try to consume gas in excess of the gas limit, an Out-of-Gas exception is thrown by the EVM.
Even though such a transaction would fail, it would be recorded on-chain and any used gas will not be refunded to the sender.
Note that in addition to the per transaction gas limit there is also a \textit{block gas limit}\footnote{At the time of writing the average block gas limit was around 10,000,000 units of gas.}, which specifies the total amount of gas that may be consumed by all transactions in a block.

\point{Gas Price}
Apart from setting a gas limit, a sender will also have to specify the \textit{gas price}, which refers to the amount of Ether the sender is willing to pay per unit of gas, generally expressed in \textit{wei} (1 wei = $10^{-18}$ETH) or \textit{Gwei} (1 Gwei = $10^{-9}$ETH).
Miners set a cut-off gas price to choose which transactions to include in their memory pool.
When constructing a new block, they then choose the transactions with the most lucrative gas prices from their memory pool.
A higher gas price will increase the fee which miners receive from a transaction, thereby motivating a miner to include a transaction in a block. 
The total amount of wei to be paid by a sender is referred to as the \textit{transaction fee} and amounts to the product of the gas price and gas cost.

\point{Gas Price Oracles}
The sender of an Ethereum transaction is exposed to the non-trivial task of having to decide on a gas price.
Since a higher gas price will increase the likelihood of having a transaction included quickly, there is a clear trade-off between waiting and paying.
We define the optimal gas price as the minimum gas price such that the transaction is included in a block within the period of time that the sender of the transaction is prepared to wait for.

In order to avoid risks of overpaying, \textit{gas price oracles} exist \cite{ethgasstation,github:geth,github:parity,github:gaspriceexpress}.
These oracles aim to recommend the gas price a transaction requires in order to be included in a block within a specified time period.
Commonly, the recommendation mechanism uses some rule-based approach analyzing the gas prices of previous blocks.
We provide a more detailed summary on existing approaches in Section~\ref{ref:related_work}.

% --- Empirical Analysis ---
\section{Empirical Analysis}
\label{ref:empirical-analysis}
\begin{figure}[h]
\centering
\includegraphics[width=\textwidth]{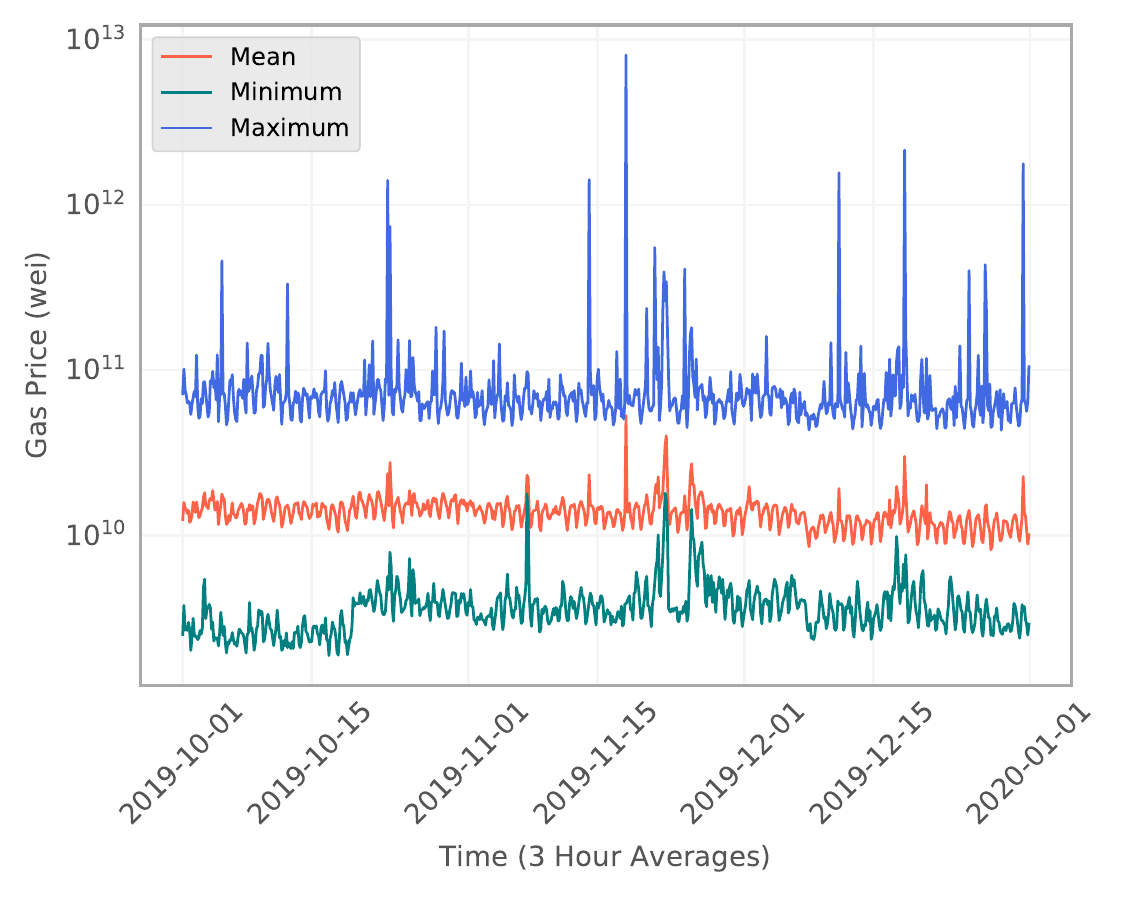}
\caption{The mean, maximum and minimum gas price averaged over 3 hour intervals from block 8,653,173 (1 October, 2019) to 9,193,265 (31 December, 2019).}
\label{fig:price_spread}
\end{figure}

In this section, we empirically analyze Ethereum block data to develop a better understanding of the gas price behavior.
We use data from the period of 1 October, 2019 to 31 December, 2019, which amounts to a total of 522,213 blocks.
When comparing mean, minimum and maximum gas prices averaged over 3 hour intervals during this period, we can see in Figure~\ref{fig:price_spread} that substantial spreads exists in the gas price.
More specifically, the maximum gas price exceeds the minimum gas price by an order of magnitude for the entire period.
% Ethereum block stats for period 1 October to 31 December 2019
\begin{table}[]
\setlength{\tabcolsep}{7pt}
\centering
\begin{tabular}{lr}
    \toprule
    Number of blocks: &  522,213 \\
    Mean gas price: & 13.9598 Gwei\\
    Median of average gas price: & 10.3260 Gwei\\
    Standard deviation of average gas price: & 46.4645 Gwei\\
    Mean gas utilization: &  79.36\% \\
    Standard deviation gas utilization: &  32.00\% \\
    \bottomrule
\end{tabular}
\caption{Mean, median and standard deviation of average gas price per block, as well as mean and standard deviation of gas utilization per block from block 8,653,173 (1 October, 2019) to 9,193,265 (31 December, 2019).}
\label{tab:empirical-gas-data}
\end{table}
The gas price volatility throughout the examined 522,213 blocks is further indicated by the standard deviation of the average gas price, which is 46.4645 Gwei at an average gas price of 13.9598 Gwei, as shown in Table~\ref{tab:empirical-gas-data}.
The same can be said about the average gas utilization per block.
\begin{figure}
\centering
\includegraphics[width=0.8\textwidth]{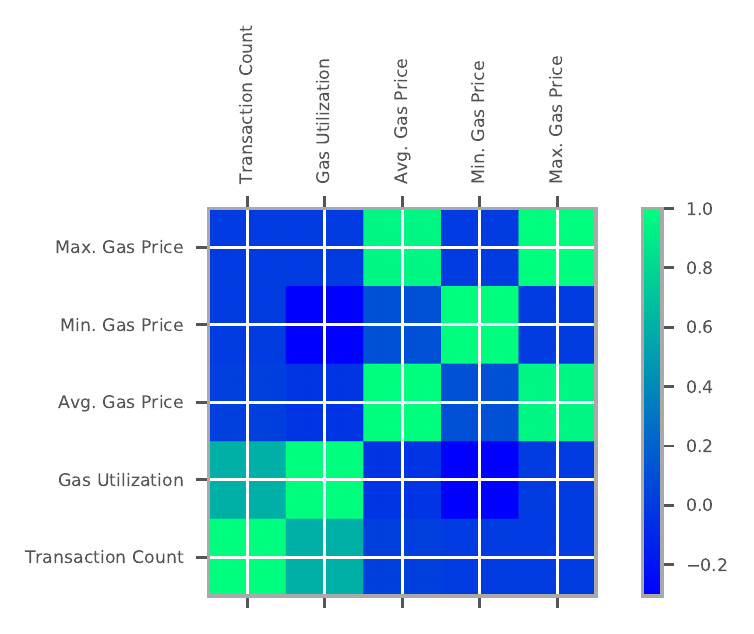}
\caption{Correlation matrix for the average gas price, maximum gas price, minimum gas price, number of transactions and gas utilization per block.}
\label{fig:corr_matrix}
\end{figure}
Figure~\ref{fig:corr_matrix} shows the cross-correlations between the average gas price, maximum gas price, minimum gas price, number of transactions and gas utilization per block.
Surprisingly, the average gas price and utilization are not correlated.
In fact, the average gas price is only significantly correlated with the maximum gas price.
The gas utilization is only correlated with the transaction count.
However, apart from these two correlated pairs, the remainder of the variables are not significantly correlated.

\begin{figure}
\centering
\includegraphics[width=1\textwidth]{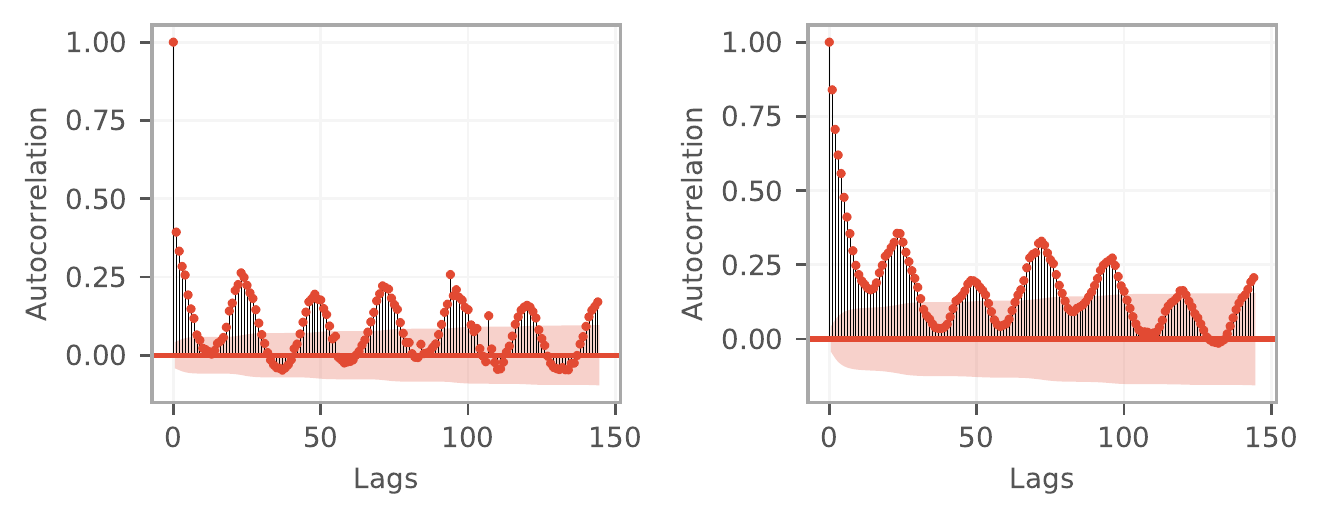}
\caption{Autocorrelation function (ACF) plot of mean (left hand side) and minimum (right hand side) gas prices averaged over one hour periods for 144 lags.}
\label{fig:acf_avg_gas_price}
\end{figure}

To investigate the presence of seasonality in the data, we examine the autocorrelation of each variable on a per block and hourly basis. 
Most interestingly, we find that even though the gas price does not exhibit any significant seasonality on a per block basis, there does exist seasonality when looking at the gas price averaged over one hour intervals, as indicated by the autocorrelation in the left plot of Figure~\ref{fig:acf_avg_gas_price}.

It can be seen that especially for a lag of 24 hours significant seasonality can be found in the data, which could be linked to different time zones of the countries where most transactions are conducted.
This seasonality can be found to an even greater extent in the autocorrelation of the minimum gas price averaged over one hour intervals.
The presence of seasonal patterns in the data alludes to the viability of machine learning models for predicting future gas prices.

% --- Methodology
\section{Methodology}
\label{ref:methodology}
The gas price recommendation methodology we propose consists of two key components.
First, we present a deep-learning based model to predict the gas price for a pre-defined period of time.
Second, we introduce an algorithm that uses these predictions to recommend a gas price for a transaction, parameterized by the sender's willingness to delay the transaction.
Both components, as well as the employed data pre-processing steps are presented in this section.

\subsection{Gas Price Prediction}
\label{methodology:prediction}
The methodology we propose requires a forecasting model to predict the gas price trajectory over a pre-defined number of time steps $s$.
In particular, we are interested in predicting the minimum gas price under rational miner behavior, since this can be seen as a lower bound for setting the gas price for a given transaction.
From the preliminary data exploration in Section~\ref{ref:empirical-analysis}, it is apparent that the per-block data is extremely noisy, which can be attenuated by averaging over a longer period of time.
We therefore average the minimum gas price of all blocks in consecutive 5 minute intervals and forecast on this level of granularity, instead of using per-block data directly.
A time step is then defined as a 5 minute interval.
We denote the complete time series of average minimum gas prices by $y$.
Furthermore, we define the aggregated time series of all features used as model input as $\mathcal{D}$, where $d_t \in \mathcal{D}$ denotes the feature vector for a single time step $t$.
For both model training and inference, we use a sliding window model that uses a fixed-size window of historical data with $l$ time steps for prediction.
The problem of forecasting a window of $s$ time steps using a window of size $l$ can then be defined as
\begin{equation}
    \hat{y}_{t+1}, \ldots , \hat{y}_{t+s} = \argmax_{y_{t+1}, \ldots , y_{t+s}} p(y_{t+1}, \ldots , y_{t+s} | d_{t-l}, ... , d_t) \,.
    \label{eq:problem_formulation_time}
\end{equation}
In the remainder of this section we present our pre-processing methodology and proposed forecasting model.

\subsubsection{Pre-processing}
\label{methodology:pre_processing}
We introduce a number of pre-processing steps to the data, which specifically aim to reduce the impact of noise while still capturing seasonal components and trends.
\begin{table}[tb]
    \centering
    \begin{tabular}{l r}
      \toprule
        \textbf{Feature Name} & \textbf{Lagged by 24h} \\
        \midrule
        Average gas price per block & Yes \\
        Transaction count per block & No \\
        Max. gas price per block & No \\
        Min. gas price per block & No \\
        ETH price at block timestamp & No \\
        \bottomrule
    \end{tabular}
    \caption{Features used as input data for the predictive model to forecast the minimum gas price.
    Lagged variables are included both with and without lag.}
    \label{tab:features}
\end{table}
Table \ref{tab:features} lists the features used as input for the predictive model.
Due to the daily seasonality in the data, some variables are also included with a lag of 24 hours.
To reduce the impact of noise in the data, we first remove outliers using a heuristic criterion, where we delete all data points that are more than 1.5 standard deviations higher or lower than the mean.
Subsequently, all data is normalized to values between 0 and 1.
Since the main goal of the predictive model is to capture the seasonality and predict the gas price on a fairly coarse level, we employ a further pre-processing step presented in \cite{DBLP:conf/icccn/Pritz0L20}.
This additional step applies a discrete Fourier transform to each window in the input data and truncates the frequency domain representation of the time series using an adaptive energy-based criterion.
We then convert it back to the time-domain using an inverse Fourier transform.
This methodology allows us to adaptively reduce the impact of short-term fluctuations in each window of input data, while still capturing the seasonal components and overall trend.

\subsubsection{Model}
\label{methodology:model}
As a forecasting model, we propose the use of a Gated Recurrent Unit (GRU) \cite{GRU_paper_2015}.
GRUs are a specialisation of recurrent neural networks, where a computationally efficient gating mechanism is used.
Gating has been shown to improve the network's ability to learn longer term dependencies \cite{Hochreiter_LSTM_1997}, making this kind of model well-suited to the problem at hand.
The GRU architecture is given by
\begin{align}
    z_t &= \sigma (W_z d_t + V_z h_{t-1} + b_z) \,, \\
    r_t &= \sigma (W_r d_t + V_r h_{t-1} + b_r) \,, \\
    h_t &= z_t \circ h_{t-1} + (1-z_t) \circ \phi (W_h d_t + V_h (r_t \circ h_{t-1}) + b_h) \,, \\
    \hat{y} &= \hat{y}_{t+1}, \ldots , \hat{y}_{t+s} = f(h_t) \,,
\end{align}
where $\circ$ denotes the Hadamard product, $W$, $V$ and $b$ are parameter matrices and biases, $\sigma(\cdot)$ and $\phi(\cdot)$ denote the sigmoid and hyperbolic tangent functions, respectively, $z_t$, $r_t$ and $h_t$ are the update and reset gates and the hidden state and $f(\cdot)$ denotes the final linear layer of the network.
The network is trained using gradient descent and backpropagation with an Adam optimiser \cite{adam_opt}.

\subsection{Recommendation Algorithm}
\label{methodology:recommendation}
We now describe our recommendation algorithm which leverages the gas prices predicted by our model. We use the 20th percentile of the predicted prices as the initial gas price, which we note $\hat{g}$. 
One of the main objectives of our algorithm is to scale $\hat{g}$ such that the faster the predicted gas prices are decreasing, the lower the gas price recommended by the algorithm.
On the other hand, if the prices are increasing, the predicted prices should not be significantly lower than the current gas price.
We incorporate this objective by finding a coefficient $0 < c \leq 1$ that is multiplied with the predicted gas price $\hat{g}$.
Furthermore, we want $c$ to increase or decrease exponentially with respect to the trend to achieve aggressive gas pricing if the predicted prices decrease quickly.

First, we compute the trend of the predictions $\hat{y}$ returned by our forecasting model.
We fit a linear function such that $\hat{y} = aX + b$, with $X = 1, 2, \cdots, s$, and store the slope $a$, which captures the trend in the predicted gas prices.
We then normalize $a$ to $\tilde{a}$ to lie in the range between $0$ to $1$.
This is achieved by computing the maximum $A_{max}$ and minimum $A_{min}$ values of the slopes we obtain for our training data and computing $\tilde{a}$ according to Equation~\eqref{eq:minmax}.
\begin{equation}
  \tilde{a} = \frac{a - A_{min}}{A_{max} - A_{min}}
  \label{eq:minmax}
\end{equation}

Then, to obtain the described exponential behavior, we exploit the fact that the exponential function in the interval $[-2, 0]$ has the desired properties and hence, compute $c$ using Equation~\eqref{eq:coefficient}.
\begin{equation}
  c = e^{2\tilde{a} - 2}
  \label{eq:coefficient}
\end{equation}

Finally, to allow the user to configure the urgency of a transaction, we define an urgency parameter $\mathcal{U}$, which we use to scale the obtained coefficient $c$ to arrive at a recommended gas price $\mathcal{G}$ given by
\begin{equation}
  \mathcal{G} = \hat{g} \cdot c \cdot \mathcal{U} \,.
  \label{eq:gas-price}
\end{equation}

\begin{algorithm}
  \begin{algorithmic}
    \Function{\texttt{EvaluateRecommender}}{\texttt{StartBlock}, \texttt{EndBlock}, \texttt{Recommend}}
      \State \texttt{Pending} $\gets \varnothing$
      \State \texttt{Results} $\gets \varnothing$
      \State \texttt{Block} $\gets$ \texttt{StartBlock}
      \While{\texttt{Block} $\leq$ \texttt{EndBlock} $\lor$ (\texttt{Pending} $\neq \varnothing \land \text{\texttt{Block}} \leq \text{\texttt{LastBlock}})$}
        \State \texttt{Price} $\gets$ \texttt{GetMinimumPrice}(\texttt{Block})
        \While{\texttt{Pending} $\neq \varnothing \land \min\limits_{t\in \text{\texttt{Pending}}}(t_1) \geq \text{\texttt{Price}}$} \Comment{$t_1$ is the transaction price}
          \State \texttt{Transaction} $\gets \argmin\limits_{t\in \text{\texttt{Pending}}}(t_1)$
          \State \texttt{Pending} $\gets$ \texttt{Pending} $\backslash$ \{\texttt{Transaction}\}
          \State \texttt{Results} $\gets \text{\texttt{Results}} \cup \{ (\text{\texttt{Transaction}}, \text{\texttt{Block}}, \text{\texttt{Price}}) \}$
        \EndWhile

        \If{\texttt{Block} $\leq$ \texttt{EndBlock}}
          \State \texttt{Recommended} $\gets$ \texttt{Recommend}(\texttt{Block})
          \State \texttt{Pending} $\gets$ \texttt{Pending} $\cup$ \{(\texttt{Block}, \texttt{Recommended})\}
        \EndIf

        \State \texttt{Block} $\gets \text{\texttt{Block}} + 1$
      \EndWhile
      \State \textbf{return} \texttt{Results}
    \EndFunction
  \end{algorithmic}
  \caption{Evaluation procedure of the gas recommendation efficiency}
  \label{alg:evaluation}
\end{algorithm}

\subsection{Measuring Gas Recommendation Efficiency}
Up to here, we have described how we recommend a price at a given block number. 
However, to understand how optimal a gas price is, we need to measure the difference between the recommended and the optimal gas price.

To evaluate the efficiency of our approach, we iterate over a range of blocks, where we do the following.
For each block, a new transaction using the recommended gas price is added to a set of pending transactions.
Each transaction in the pending set is processed upon encountering a block with a minimum gas price lower than that specified in the transaction.
We keep track of the recommended price, the inclusion price, i.e. the minimum gas price of the block where the transaction is included, and the number of blocks elapsed until inclusion.
We show the detailed steps in Algorithm~\ref{alg:evaluation}.
The \texttt{EvaluateRecommender} function takes a start block, an end block and a recommendation function to evaluate.
\texttt{LastBlock} is the number of the last block which we evaluate and \texttt{GetMinimumPrice} returns the minimum gas price for a given block.

To be able to evaluate the efficiency of our algorithm, we use the Geth gas price recommendation algorithm as the main baseline, as it is by far the most widely used Ethereum client~\cite{ethernodes}.

% --- Results ---
\section{Results}
\label{ref:results}
In this section, we present the results we obtain when using the methodology presented in the previous section and compare them with our baselines.
\begin{figure}
  \includegraphics[width=\textwidth]{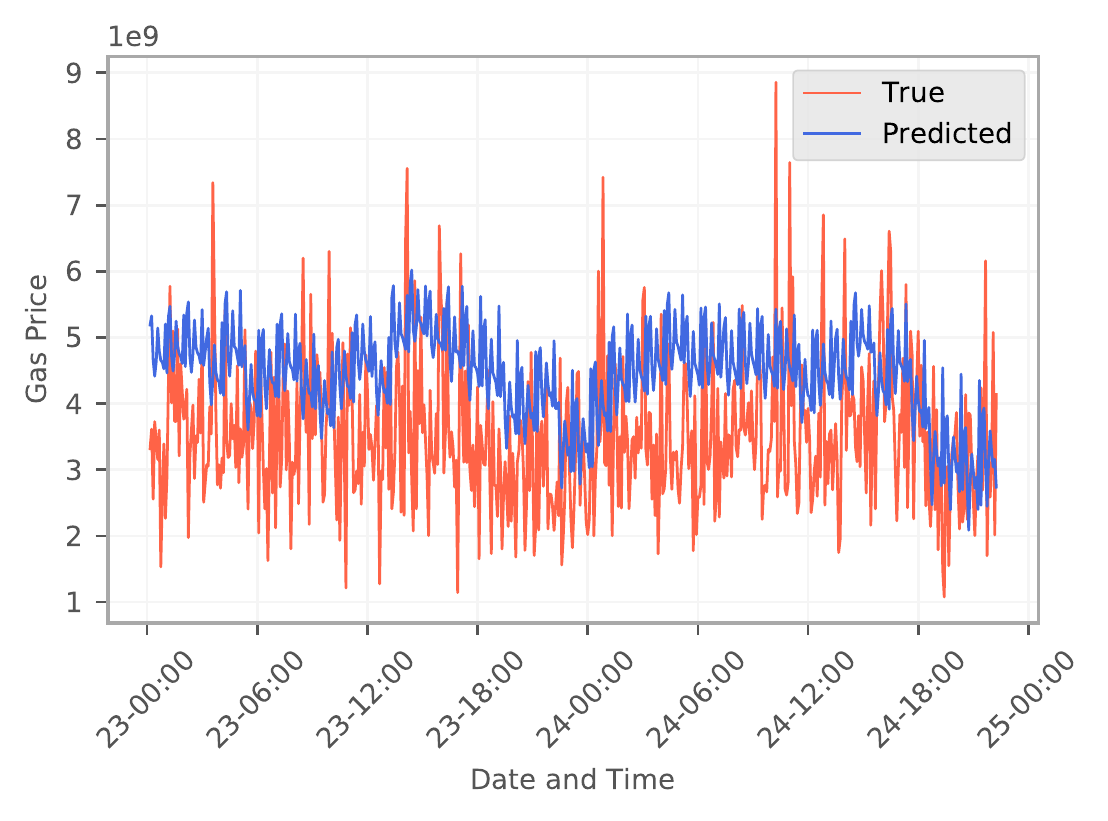}
  \caption{Exemplary gas price predictions obtained with our forecasting model for the period between the 23 November, 2019 and 25 November, 2019.}
  \label{fig:gas-price-predictions}
\end{figure}

\subsection{Model Training}
\label{results:model_training}
All models are implemented in Python, using the PyToch library \cite{pytorch_docs}.
We train all models on a personal computer with 32GB of RAM, an $8^{th}$ generation Intel Core i7-8700 with 3.20GHz and 6 cores and a 256GB SATA hard drive.
Model training and hyper parameter tuning is performed on the data between 10 November, 2019 to 20 November, 2019, where we use the first 70\% of the data for training and the remaining 30\% for validation. 
We show exemplary predictions of our model in Figure~\ref{fig:gas-price-predictions}.

\subsection{Evaluation}
\label{results:evaluation}
We use a sample of around five days of data --- from 20 November, 2019 (block 8,965,759) to November 24, 2019 (block 8,995,344) --- and evaluate the different price recommendation strategies using the procedure presented in Algorithm~\ref{alg:evaluation}.
\begin{table}
  \centering
  \setlength{\tabcolsep}{3pt}
  \begin{tabular}{l l l}
    \toprule
    \textbf{Model} & \textbf{Parameter} & \textbf{Description}\\
    \midrule
    \multirow{2}{*}{Geth} & \multirow{2}{*}{Scaling ($\mathcal{S}$)} & Ratio by which to scale the price\\
        &               &  (0.8 means use 80\% of the recommended price)\\
    proposed approach & Urgency ($\mathcal{U}$) & Urgency tuning parameter to trade-off price for time\\
    Look-ahead & Blocks ($\mathcal{B}$) & Number of blocks to look ahead\\
    \bottomrule
  \end{tabular}
  \caption{Parameters used in the different strategies}
  \label{tab:parameters}
\end{table}
We first describe the parameters of each strategy in Table~\ref{tab:parameters}. 
For Geth we use a scaling ratio parameter $\mathcal{S}$ with which the recommended gas price is multiplied. 
The main purpose of this parameter is to ensure that giving a lower gas price does have a direct impact on the number of blocks waited.
Our proposed recommendation strategy accepts a single parameter $\mathcal{U}$ representing the urgency. 
The urgency parameter is used to trade off gas price for waiting time: the lower the urgency, the lower the gas price and hence, the longer the waiting time. 
Empirically, reasonable values for these parameters are roughly between $0.7$ and $1.3$, where $0.7$ will result in cheap but long to be accepted transactions and $1.3$ will result in more expensive but faster transactions. 
Finally, our look-ahead model, which we use to estimate the lowest possible price takes a parameter $\mathcal{B}$ representing the maximum look-ahead as a number of blocks. 
We note that the look-ahead strategy is for validation purposes only as it uses information about future blocks, which would obviously not be available in practice.

\begin{table}
  \centering
  \setlength{\tabcolsep}{5pt}
  \begin{tabular}{l l r r}
    \toprule
    \textbf{Strategy} & \textbf{Parameter} & \textbf{Gas price} & \textbf{Blocks waited}\\
    \midrule
    Geth & $\mathcal{S} = 1.0$ & 4,414,902,746 & 1.97\\
    Geth & $\mathcal{S} = 0.9$ & 4,080,968,868 & 15.49\\
    Geth & $\mathcal{S} = 0.8$ & 3,531,922,197 & 25.52\\
    \midrule
    Look-ahead & $\mathcal{B} = 15$ & 1,166,965,099 & 4.80\\
    Look-ahead & $\mathcal{B} = 30$ & 969,559,938 & 8.52\\
    Look-ahead & $\mathcal{B} = 60$ & 782,105,012 & 18.84\\
    \midrule
    Proposed approach & $\mathcal{U} = 1.0$ & 2,120,108,703 & 3.28\\
    Proposed approach & $\mathcal{U} = 0.9$ & 1,908,097,833 & 3.79\\
    Proposed approach & $\mathcal{U} = 0.8$ & 1,696,086,963 & 5.13\\
    Proposed approach & $\mathcal{U} = 0.7$ & 1,484,076,092 & 10.06\\
    \bottomrule
  \end{tabular}
  \caption{Results of the different recommendation strategies presented. Gas price and wait time are averaged over the number of blocks processed. Parameters are described in Table~\ref{tab:parameters}.}
  \label{tab:results}
\end{table}
We present a summary of the results for the different recommendation strategies in Table~\ref{tab:results}. 
We use several values for the parameters of each strategy and order its results so that the gas price decreases and the number of blocks to wait increases. 
We can see that using the price recommended by Geth, the waiting time is very short --- on average less than 2 blocks --- with an average gas price of around~4.4 Gwei. 
However, by just using 90\% of the recommended price, the waiting time increases to an average of 15.5 blocks. 
Comparing these results to the minimum possible gas price, obtained from the look-ahead model, we can see that by only waiting for an average of 4.8 blocks a saving of 75\% could be obtained.
Although these numbers are hypothetical, they suggest the potential for significant improvement.

\begin{figure}
  \includegraphics[width=\textwidth]{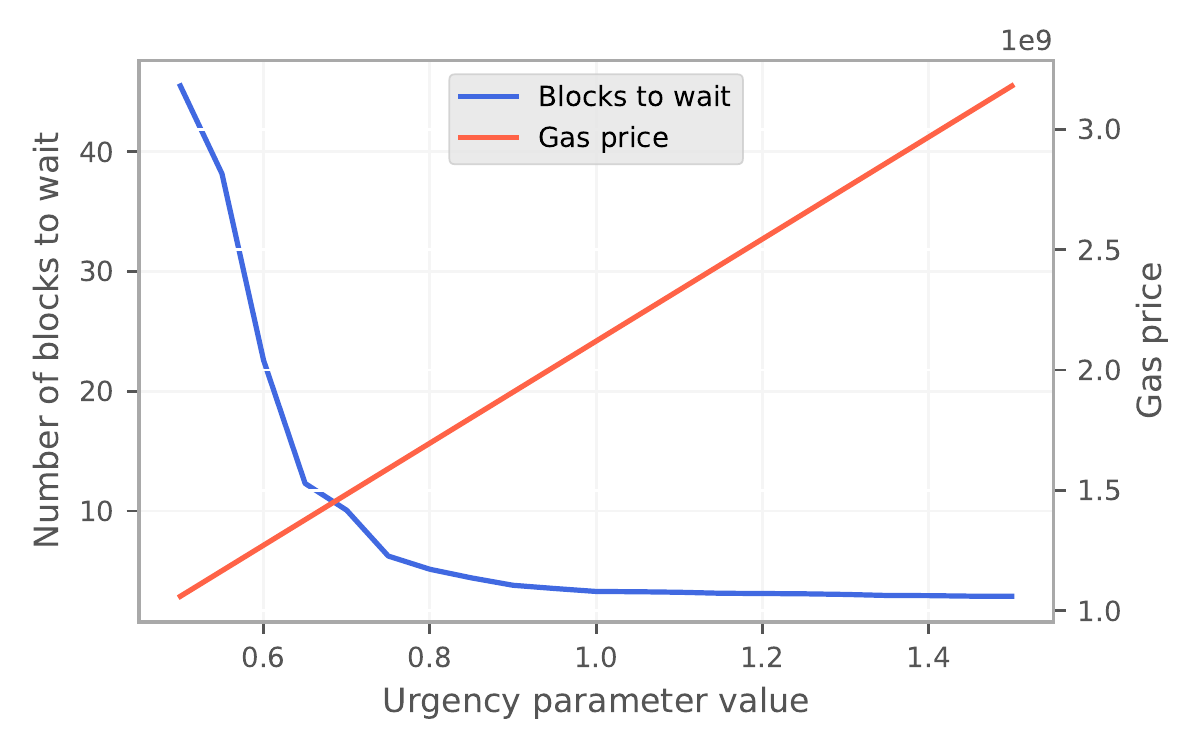}
  \caption{Effect of the urgency parameter on the average gas price paid and number of blocks waited.}
  \label{fig:urgency-effect}
\end{figure}

We now show how our model performs in comparison to the price recommended by Geth and the hypothetical minimum price. 
With the urgency parameter set to $1.0$, our model recommends a gas price on average twice as low as the Geth price, while waiting for an average of approximately 3.3 blocks. 
When decreasing the urgency parameter, we can see that the number of blocks elapsed increases fairly slowly at first but doubles between 0.8 and 0.7, showing that at this point the gas price becomes too low for the transaction to be included in a timely manner. 
In Figure~\ref{fig:urgency-effect}, we show the effect of our urgency parameter on the average gas price paid and the average number of blocks elapsed until the transaction is included.

% --- Related Work ---
\section{Related Work}
\label{ref:related_work}
For Ethereum in particular, there has been extensive research on smart contract correctness, upper-bound gas consumption and imperfections in the current EVM gas cost model.
Nonetheless, very little work has been done with the goal of determining optimal gas prices.
In this section, we first present existing work on the gas mechanism, before examining the most widely used gas price recommendation methods.

\subsection{Gas Mechanism}
The overconsumption of gas can be harmful for the contract user for two main reasons: higher monetary costs and potential vulnerabilities.
Gas overconsumption is examined by Chen et al.~\cite{Chen2017}, who focus on gas usage optimization by introducing Gasper, a tool leveraging symbolic execution for detecting costly patterns in the bytecode of smart contracts which are not optimized by the Solidity compiler.
Potential issues in the form of gas-related vulnerabilities are carefully examined by Grech et al.~\cite{Grech2018}, who propose a static analysis tool, called MadMax, predominantly suitable for detecting out-of-gas exceptions which may cause contract funds being locked.
Elvira et al.~\cite{elvira2018gastap} present Gastap, a static analysis tool for inferring gas upper bounds for smart contracts and are thereby able to detect whether any out-of-gas vulnerabilities could exist.
A further approach for computing gas consumption upper bounds was introduced by Marescotti et al.~\cite{marescotti2018computing}, however, the authors are yet to implement and test their algorithms in an EVM setting.
For a more general summary of existing smart contract verification tools we point the reader to \cite{harz2018towards}.

There have been several pieces of work focusing on imperfections in the current gas cost mechanism.
Both Yang et al.~\cite{yang2019empirically} and Perez and Livshits~\cite{perez2019broken} identify inconsistencies in the pricing of EVM instructions in the current gas cost model.
The latter propose a new type of attack aimed at exploiting EVM design flaws by generating resource exhaustive contracts, which are on average significantly slower in terms of throughput than typical contracts.
As a means of preventing Denial-of-Service attacks stemming from under-priced EVM instructions a modification of the current gas cost mechanism has been proposed by~\cite{Chen2017Metering}.

While several pieces of existing work examine the current gas cost mechanism, limited work exists on gas price recommendation.
Pierro et al.~\cite{pierro2019influence} investigate potential factors that influence transaction fees in Ethereum from a technical and economic perspective, yet leave a gas price prediction model for future research. 

\subsection{Gas Price Oracles}
\label{ref:gas_prediction_oracles}
In the following, we examine existing approaches for gas price recommendation that are used in practice.

\point{Geth}
The Ethereum client implementation in go, namely Geth~\cite{github:geth}, accounts for over 79\% of all Ethereum clients~\cite{ethernodes}. To recommend a gas price, Geth uses the minimum gas price of the previous blocks. It looks back at the 100 blocks preceding the current one and then uses the value of the 60th percentile of the minimum gas prices as the price recommendation.

\point{EthGasStation}
A further gas price oracle has been introduced by EthGasStation~\cite{ethgasstation}, a third-party tool, which estimates the expected number of blocks required to confirm a transaction at a given gas price using a Poisson regression model based on data of the previous 10,000 blocks.
This approach has also been implemented by the popular Ethereum block explorer Etherchain~\cite{etherchain}. 
Unfortunately, no historical data was available for comparison.

\point{GasStation -- Express}
EthGasStation also released a more simple gas price oracle called ``GasStation -- Express"~\cite{github:gaspriceexpress}.
This approach predicts the likelihood of a transaction being included in the next block at a given gas price by examining the percentage of the last 200 blocks that included a transaction with the same or lower gas price~\cite{medium:gaspriceexpress}.
The percentage thresholds of recent block inclusions are fixed for the categories \textit{Fast} (90\%), \textit{Standard} (60\%) and \textit{SafeLow} (35\%).
Additionally a \textit{Fastest} option is given, whereby the suggested gas price was included by all of the previous 200 mined blocks, which likely results in the sender overpaying considerably.
Just like the threshold percentages, the associated expected confirmation times are also hard-coded, which limits the speed at which the system can react to changes.

% --- Conclusion ---
\section{Conclusion}
\label{ref:conclusion}
Motivated by an empirical analysis of 3 months of data, we have proposed a novel approach for recommending the Ethereum gas price that outperforms the method of the most widely used Ethereum client.
Our approach uses a deep-learning based price forecasting model as well as an algorithm parameterized by an urgency value that can be set by the user.
In a comprehensive evaluation, we show that our approach is able to reduce the average gas price paid by the sender of a transaction by more than 50\% while only introducing an average additional waiting time of 1.3 blocks compared to Geth.

Our evaluation of the proposed approach aimed to focus on common-sized transactions.
For more computationally intensive transactions, the gas price would likely need to be increased to ensure timely inclusion in a block.
However, this could be easily accomplished by adjusting the urgency parameter.

Future work can examine the usefulness of additional data, such as memory pool data, as model inputs.
Additionally, the evaluation and comparison of our approach and previous approaches in a larger simulation may be a fruitful avenue for further research.

\bibliographystyle{abbrv}
\bibliography{references.bib}

\end{document}